\documentclass{appolb}
\usepackage{amsmath}
\usepackage{amssymb}
\usepackage{cite}
\usepackage{graphicx}
\usepackage[symbol]{footmisc}
\usepackage{url}
\usepackage{slashed}
\usepackage[square,numbers,sectionbib,compress]{natbib}
\usepackage{graphicx}
\usepackage{breqn}
\usepackage{bbold}
\usepackage{xcolor}
\usepackage{float}
\usepackage[colorlinks=true,urlcolor=black,anchorcolor=black,citecolor=black,filecolor=black,linkcolor=black,menucolor=black,linktocpage=true,pdfproducer=medialab,pdfa=true]{hyperref}
\allowdisplaybreaks

\newcommand{\eps}{\varepsilon}

\begin{document}
%\begin{titlepage}
%
\begin{center}
{\LARGE
    Anomalous dimensions for exclusive processes\footnote[1]{Presented at the V4-HEP 2 - Theory and Experiment in High Energy Physics Workshop, Budapest, Hungary, 12-14 March 2024.}
}
\vspace{0.5cm}

\large
S.~Van Thurenhout$^{\, a}$\footnote[2]{Speaker}\\
\vspace{0,5cm}
\small
{$^{\, a}$HUN-REN Wigner Research Centre for Physics, Konkoly-Thege Mikl\'os u. 29-33, 1121 Budapest, Hungary}
\end{center}
%\begin{center}
%    {\small\itshape (Received \today)}
%\end{center}
%
We give an overview of recent developments in the computation of the anomalous dimension matrix of composite operators in non-forward kinematics. The elements of this matrix set the evolution of non-perturbative parton distributions such as the generalized parton distribution functions. The latter provide important information about hadronic structure and are accessible experimentally in hard exclusive scattering processes. We focus our discussion on a recent method that exploits consistency relations for the anomalous dimensions which follow from the renormalization structure of quark and gluon operators.
%
%\vspace*{0.3cm}
%\end{titlepage}

\section{Introduction}
\label{sec:intro}
Understanding the structure of hadrons is an important topic of modern quantum chromodynamics (QCD). This structure is described by non-perturbative parton distributions, which correspond to hadronic matrix elements of composite quark and gluon operators. Their scale dependence, which is important input for phenomenological studies, is characterized by the perturbative anomalous dimensions of these operators. In non-forward kinematics, relevant for exclusive processes such as deeply-virtual Compton scattering, the operators mix under renormalization with total-derivative operators which complicates the extraction of the anomalous dimensions. Two methods to help in this extraction are reviewed in this work.

\section{Operator anomalous dimensions}
\label{sec:operators}
The anomalous dimensions of interest are computed by renormalizing the \textit{partonic} matrix elements of the operators that define the parton distributions. In this work we focus on leading-twist operators\footnote[3]{For a recent overview of higher-twist corrections see \cite{Braun:2022gzl}.} of spin $N$. Concentrating on unpolarized processes, the following operators contribute
\begin{align}
\label{eq:Ons}
        &\qquad\qquad\mathcal{O}^{q,\text{NS}}_N = \overline{\psi}\lambda^{\alpha}\slashed\Delta D^{N-1}\psi\,,\\&
        \mathcal{O}^{q,\text{S}}_N = \overline{\psi}\slashed\Delta D^{N-1}\psi\,, \quad \mathcal{O}^{g,\text{S}}_N = F_{\mu \Delta}D^{N-2}F_{\Delta}{}^{\mu}\,.
\label{eq:Os}
\end{align}
Here $D_{\mu} = \partial_{\mu}-i g_s A_{\mu}$ is the QCD covariant derivative, $F_{\mu\nu}$ is the gluon field strength and $\Delta$ is an arbitrary lightlike vector. We also introduced the shorthands $D\equiv\Delta^{\mu}D_{\mu}$ and $F_{\mu\Delta}\equiv \Delta^{\nu}F_{\mu\nu}$. The operator in Eq.~(\ref{eq:Ons}) is a flavor-non-singlet as signalled by the appearance of the flavor generator $\lambda^{\alpha}$. The operators in Eq.~(\ref{eq:Os}) on the other hand are flavor-singlet and will mix under renormalization. As we are interested in non-forward distributions, we need to take into account mixing with total-derivative operators. For example, for $\mathcal{O}^{q,\text{NS}}_N$ the renormalization takes on the form
{\begin{equation}
        \begin{pmatrix}
            \mathcal{O}^{q,\text{NS}}_{N+1}\vspace{2pt} 
            \\ \partial \mathcal{O}^{q,\text{NS}}_{N}\\ \vdots \\ \partial^N \mathcal{O}^{q,\text{NS}}_{1}
    \end{pmatrix} 
    \,=\, 
    \begin{pmatrix}
            {Z_{N,N}^{qq,\text{NS}}} &  Z_{N,N-1}^{qq,\text{NS}} & ... &  Z_{N,0}^{qq,\text{NS}} \vspace{2pt}\\
            0 & {Z_{N-1,N-1}^{qq,\text{NS}}} & ... &  Z_{N-1,0}^{qq,\text{NS}} \\
            \vdots & \vdots & ... & \vdots  \\
            0 & 0 & ...  &  {Z_{0,0}^{qq,\text{NS}}}
    \end{pmatrix} 
    \begin{pmatrix}
            [\mathcal{O}^{q,\text{NS}}_{N+1}] \vspace{2pt}\\ [\partial \mathcal{O}^{q,\text{NS}}_{N}] \\ \vdots 
             \\ [\partial^N \mathcal{O}^{q,\text{NS}}_{1}]
    \end{pmatrix}
    \end{equation}}
in which square brackets denote the renormalized operators. The anomalous dimension matrix (ADM) is then defined as
    {\begin{equation}
        \hat{\gamma}^{qq,\text{NS}} =-\frac{\text{d}\ln \hat{Z}^{qq,\text{NS}}}{\text{d}\ln\mu^2} = \begin{pmatrix}
            {\gamma^{qq,\text{NS}}_{N,N}} & \gamma^{qq,\text{NS}}_{N,N-1} & ... & \gamma^{qq,\text{NS}}_{N,0}\vspace{2pt} \\
            0 & {\gamma^{qq,\text{NS}}_{N-1,N-1}} & ... & \gamma^{qq,\text{NS}}_{N-1,0} \\
            \vdots & \vdots & ... & \vdots  \\
            0 & 0 & ...  &  {\gamma^{qq,\text{NS}}_{0,0}}
    \end{pmatrix} .
    \end{equation}}
The diagonal elements of the ADM correspond to the forward anomalous dimensions, which determine the scale dependence of the forward parton distributions. We now briefly review two methods to reconstruct the off-diagonal elements.\footnote{Another method, which is independent of the ones discussed here, can be found in \cite{Kisselev:2005er,Kiselev:2011zc}.} The first is based on conformal symmetry arguments while the second uses consistency relations following from the renormalization structure of the operators.

\subsection{Anomalous dimensions from conformal symmetry}
This approach starts from QCD in $D=4-2\eps$ dimensions at the critical point, which is characterized by a vanishing beta-function. In this regime, the conformal algebra can be used to write down relations between the operator anomalous dimensions, allowing the reconstruction of the full ADM, $\hat{\gamma}^{\mathcal{C}}$. The physical anomalous dimensions are then equivalent to the conformal ones, up to terms proportional to the (non-zero) QCD beta-function and the so-called conformal anomaly. The latter is a perturbative quantity and is currently known to two loops \cite{Mueller:1991gd,Braun:2016qlg,Braun:2017cih,Manashov:2024fcd}, allowing for the reconstruction of the anomalous dimensions up to three loops \cite{Braun:2017cih,Manashov:2024fcd}. A particularly nice feature of this method is that the one-loop off-diagonal elements of the ADM vanish due to exact conformal symmetry at leading order, while the main bottleneck is the determination of the anomaly.

\subsection{Anomalous dimensions from consistency relations}
A second approach to determine the off-diagonal elements of the ADM is the use of relations between total-derivative operators. In fact, writing\footnote{We omit the S and NS superscripts here. Furthermore, the relation for the quark operator trivially generalizes to the singlet case by simply discarding $\lambda^{\alpha}$.}
\begin{equation}
    \mathcal{O}^{q}_{r,s,t} = 
    \partial^{r} ((D^{s}\overline{\psi}) \lambda^{\alpha}\slashed\Delta 
    (D^{t}\psi))\text{ and }\mathcal{O}^{g}_{r,s,t} = 
    \partial^{r} ((D^{s}F_{\mu\Delta})(D^{t}F_{\mu\Delta}))\,,
\end{equation}
we have
\begin{equation}
     \mathcal{O}^{q/g}_{r,s,t} = \mathcal{O}^{q/g}_{r-1,s+1,t} + \mathcal{O}^{q/g}_{r-1,s,t+1}\,.
\end{equation}
These relations, which are valid for the bare operators, can be turned into recursive identities by applying them to $\mathcal{O}^{q/g}_{N,0,0}$. Performing the renormalization then leads to non-trivial relations between the anomalous dimensions. For example, for the non-singlet quark operators one particular form of the resulting relation is \cite{Moch:2021cdq}
\begin{equation}
\label{eq:relQ}
    \forall k: \qquad 
    \sum_{j=k}^{N}\Bigg\{(-1)^k\binom{j}{k}\gamma^{qq,\text{NS}}_{N,j}-(-1)^j\binom{N}{j}\gamma^{qq,\text{NS}}_{j,k}\Bigg\} = 0
    \, .
\end{equation}
The latter is valid to all orders in perturbation theory and, as explained in \cite{Moch:2021cdq}, can be used to reconstruct the off-diagonal elements of the ADM. A nice consequence of Eq.~(\ref{eq:relQ}) is that the next-to-diagonal (NTD) elements, $\gamma^{q,\text{NS}}_{N,N-1}$, are directly related to the forward anomalous dimensions,
\begin{equation}
    \gamma^{qq,\text{NS}}_{N,N-1} = \frac{N}{2}\left(\gamma^{qq,\text{NS}}_{N-1,N-1}-\gamma^{qq,\text{NS}}_{N,N}\right)\,.
\end{equation}
Similar identities can also be written down for the flavor-singlet anomalous dimensions. Consider, e.g., the insertion of the gluon operator in a gluon two-point function. The corresponding anomalous dimensions are denoted by $\gamma^{gg}_{N,k}$, for which we find the following relation
\begin{equation}
\label{eq:relG}
    \forall k: \qquad 
    \sum_{j=k}^{N}\Bigg\{(-1)^k\binom{j-1}{k-1}\gamma^{gg}_{N,j}-(-1)^j\binom{N-1}{j-1}\gamma^{gg}_{j,k}\Bigg\} = 0\,.
\end{equation}
Note that now $k>0$, as $k=0$ would be associated to (derivatives of) a spin-one gluon operator. As before, the NTD elements are directly related to the forward anomalous dimensions,
\begin{equation}
\label{eq:NTDg}
    \gamma^{gg}_{N,N-1} = \frac{N-1}{2}\left(\gamma^{gg}_{N-1,N-1}-\gamma^{gg}_{N,N}\right)\,.
\end{equation}
In deriving Eqs.~(\ref{eq:relG}) and (\ref{eq:NTDg}), we ignored mixing with other types of operators, which is valid to one-loop accuracy. Beyond this order, mixing with the flavor-singlet quark operators and non-gauge-invariant operators needs to be taken into account. Such extensions are left for future considerations. Nevertheless, we expect the identities presented above to stay valid \textit{beyond} the one-loop level. To see this, consider the following basis transformation formula, which takes quantities from the non-conformal basis to the conformal one \cite{VanThurenhout:2023gmo}
\begin{align}
     \label{eq:basisTrans}
     \begin{split}
    \gamma_{N,k}^{gg,\mathcal{C}} &= \frac{(-1)^k(3+2k) k!}{(N-1)!}\sum_{l=k}^{N}(-1)^l\binom{N-1}{l-1}\frac{(N+l+2)!}{(l+1)!}\\&\qquad\qquad\qquad\qquad\times\sum_{j=k}^{l}\binom{j}{k}\frac{(j+1)!}{j(j+k+3)!}\gamma_{l,j}^{gg}\,.
    \end{split}
\end{align}
This type of similarity transformation is independent of the details of the operator renormalization and hence valid to all orders in perturbation theory. Now, in the conformal basis, the anomalous dimensions vanish whenever $N-k$ is odd, which is a consequence of CP-symmetry. So, setting $k=N-1$, the left-hand side of Eq.~(\ref{eq:basisTrans}) vanishes and we find
\begin{equation}
    0 = \gamma_{N-1,N-1}^{gg}-\frac{2}{N-1}\gamma_{N,N-1}^{gg}-\gamma_{N,N}^{gg}\,,
\end{equation}
exactly reproducing Eq.~(\ref{eq:NTDg}) but now as an \textit{all-order} identity. Using similar arguments for different values of $N-k$, we can then conclude that also the general relation in Eq.~(\ref{eq:relG}) remains valid to all orders in perturbation theory.

\section{Summary}
\label{sec:conclusion}
We have reviewed two methods to compute the anomalous dimensions of composite operators in non-forward kinematics using (a) conformal symmetry arguments and (b) consistency relations following the renormalization structure of the operators. Both methods come with their own advantages, and, as such, can be used in a complementary fashion.

\subsection*{Acknowledgements}
Part of this work has been supported by grant K143451 of the National Research, Development and Innovation Fund in Hungary.

{\scriptsize
\bibliographystyle{JHEP}
\bibliography{omebib}
}

\end{document}